\documentclass{ws-ijmpa}
\newcommand{\adb}{\allowdisplaybreaks }
\newcommand{\bs}{\begin{subequations}}
\newcommand{\es}{\end{subequations}}
\begin{document}

\markboth{Nail Khusnutdinov}
{Self-interaction for particles in the wormhole space-times}

%
\catchline{}{}{}{}{}
%

\title{SELF-INTERACTION FOR PARTICLES IN THE WORMHOLE SPACE-TIMES}

\author{NAIL KHUSNUTDINOV}

\address{Institute of Physics, Kazan Federal University, Kremlevskaya
18\\
Kazan, 420008, Russia\\
7nail7@gmail.com}

\maketitle

\begin{history}
\received{24 June 2011}
\end{history}

\begin{abstract}
The self-energy and self-force for particles with electric and scalar charges at
rest in the space-time of massless and massive wormholes are considered. The
particle with electric charge is always attracted to wormhole throat for
arbitrary profile of the throat. The self-force for scalar particle shows
different behavior depending on the non-minimal coupling. The self-force for
massive scalar field is localized close to the throat of the wormhole.  
\end{abstract}

\ccode{PACS numbers: 04.40.-b, 98.80.Cq}

\section{Introduction}\label{Sec:Intro}

Wormholes are topological handles linking different regions of the
Universe or different universes. The activity in wormhole physics
was first initiated by the classical paper by Einstein and Rosen,\cite{EinRos35} and later by Wheeler.\cite{WheBook} The latest
growth of interest in wormholes was connected with the ``time
machine''\/, introduced by Morris, Thorne and Yurtsever in Refs.
\refcite{MorTho88,MorThoYur88}. Their works led to a surge of activity
in wormhole physics.\cite{VisBook} The main and unsolved problem in
wormhole physics is whether wormholes exist or not. The wormhole has
to violate energy conditions and the source of the wormhole geometry
should be exotic matter. One example\cite{Kra00,KhuSus02,Gar05} of
such exotic matter is quantum fluctuations, which may violate the
energy conditions. Another possible sources are a scalar field with
reversed sign of kinetic term,\cite{Arm02} and cosmic phantom
energy.\cite{Sus05,Lob05} The question of wormhole's stability is
not simple and requires subtle calculations (see, for example\cite{Lob05_2}). The problem arises because a wormhole needs some
amount of exotic matter which violates energy conditions and has
unusual properties. Recently it was observed that some observational
features of black holes can be closely mimicked by spherically
symmetric static wormholes\cite{DamSol07} having no event horizon.
Some astronomical observations indicate possible existence of black
holes (see, for example, Ref. \refcite{Nar05}). It is therefore
important to consider possible astronomical evidences of the
wormholes. Some aspects of wormholes' astrophysics were considered in
Ref. \refcite{KarNovSha06}, where it was noted that this massive
compact object may correspond to a wormhole of macroscopic size with
strong magnetic field. A matter may go in and come back out of the
wormhole's throat.

In the framework of general relativity there exists a specific
interaction of particles with gravitating objects -- the
gravitationally induced self-interaction force which may have
considerable effect on the wormholes' physics. It is well-known\cite{DeWBre60} that in a curved background alongside with the
standard Abraham-Lorenz-Dirac self-force there exists a specific
force acting on a charged particle. This force is the
manifestation of non-local essence of the electromagnetic field.
It was considered in details in some specific space-times (see
Refs. \refcite{Poi,Khu05} for review). For example, in the case of
the straight cosmic string space-time the self-force appears to be
the only form of interaction between the particle and the string.
Cosmic string has no Newtonian potential but nevertheless a
massless charged particle is repelled by the string,\cite{Lin86} 
whereas massive uncharged particle is attracted by the string due
to the self-force.\cite{Gal90} The non-trivial internal structure
of the string does not change this conclusion.\cite{KhuBez01} The
potential barrier appears which prevents the charged particle from
penetrating into the string. For GUT cosmic strings the potential
barrier is $\sim 10^5 Gev$. The wormhole is an example of the
space-time with non-trivial topology. The consideration of the
Casimir effect for a sphere that surrounds the wormhole's throat
demonstrates an unusual behavior of the Casimir force\cite{KhaKhuSus06} -- it may change its sign depending on the
radius of the sphere. It is expected that the self-force in the
wormhole space-time will show an unusual behaviour too. 

In this paper we review the self-interaction force for particle with electric
and scalar charges in different kind of wormholes space-times. 

\section{The Massless Wormhole Space-Time}

Let us consider an asymptotically flat wormhole space-time. We
choose the line element of this space-time in the following form
\begin{equation}
ds^2 = -dt^2 + d\rho^2 +r^2(\rho)(d\theta^2 + \sin^2 \theta
d\varphi^2),\label{ds^2}
\end{equation}
where $t,\rho \in \mathbf{R}$ and
$\theta \in [0,\pi], \varphi \in [0,2\pi]$. Profile of the
wormhole throat is described by the function $r(\rho)$. This space-time
firstly was considered by Bronnikov\cite{Bron73} and Ellis.\cite{Ell73} This
function must have a minimum at $\rho = 0$, and the minimal value
at $\rho = 0$ corresponds to the radius, $a$, of the wormhole
throat, \begin{displaymath} r(0) = a,\ \dot r(0) = 0, \end{displaymath} where an
over dot
denotes the derivative with respect to the radial coordinate
$\rho$. Space-time is naturally divided into two parts in
accordance with the sign of $\rho$. We shall label the part of the
space-time with positive (negative) $\rho$ and the functions on
this part with the sign "+"\ ("--").

The space-time possesses non-zero curvature. The scalar curvature
is given by
\begin{displaymath}
R = -\frac{2(2r\ddot r + \dot r^2 -1)}{r^2}.
\end{displaymath}
Far from the wormhole throat the space-time becomes Minkowskian,
\begin{equation}
r(\rho)|_{\rho \to \pm \infty} = \pm \rho.\label{condinfin}
\end{equation}

Various kinds of throat profiles have been already considered in
another context.\cite{KhuSus02} The simplest model of a wormhole
is that with an infinitely short throat,\cite{KhuSus02}
\begin{equation}
r = a+|\rho|.
\end{equation}
The space-time is flat everywhere except for the
throat, $\rho =0$, where the curvature has delta-like form,
\begin{displaymath}
R = -8\frac{\delta (\rho)}{a}.
\end{displaymath}
Another wormhole space-time
that is characterized by the throat profile
\begin{equation}
r = \sqrt{a^2+\rho^2}
\end{equation}
is free of curvature singularities:
\begin{displaymath}
R = -\frac{2a^2}{(a^2+\rho^2)^2}.
\end{displaymath}
The wormholes with the following profiles of throat
\begin{eqnarray*}
r &=& \rho \coth \frac{\rho}b + a-b,\adb\\
r &=& \rho \tanh \frac{\rho}b + a,
\end{eqnarray*}
have a throat whose length may be described using a parameter
$b$. The point is that for $\rho > b$ the space-time becomes
Minkowskian exponentially fast. 

\section{Self-Energy and Self-Force}\label{Sec:3}

The self-energy of particle at rest with electric charge $e$ is defined as
one-half of coincidence limit of the renormalized interaction energy of
particle with the same charge. For particle at rest with scalar charge the
expression for self-energy was found in Ref. \refcite{BezKhu} and it has the
same form as for electric case 
\begin{equation}
U = \frac{e^2}{2}G^{ren}(x,x).
\end{equation}

Let us consider\cite{KhuBah} a charged particle at rest in the point
$\rho',\theta',\varphi'$ in the space-time with metric (\ref{ds^2}). The Maxwell
equation for zero component of the potential reads
\begin{equation*}
\triangle A^0 = - \frac{4\pi e \delta (\rho - \rho')}{r^2(\rho)}
\frac{\delta (\theta - \theta') \delta (\varphi -
\varphi')}{\sin\theta}
\end{equation*}
where $\triangle = g^{kl} \nabla_k\nabla_l$. Due to static
character of the background we set other components of the vector
potential to be zero. It is obvious that $A^0 = 4\pi e
G(\mathbf{x};\mathbf{x'})$, where the three-dimensional Green's
function $G$ obeys the following equation
\begin{equation*}
\triangle G(\mathbf{x};\mathbf{x'}) = - \frac{\delta (\rho -
\rho')}{r^2(\rho)} \frac{\delta (\theta - \theta') \delta (\varphi -
\varphi')}{\sin\theta}.
\end{equation*}
Due to spherical symmetry we may extract the angular dependence
(denoting succinctly $\Omega = (\theta,\varphi)$)
\begin{equation*}
G(\mathbf{x};\mathbf{x'}) = \sum_{l=0}^\infty \sum_{m=-l}^l
Y_{lm}(\Omega) Y_{lm}^*(\Omega') g_l(\rho,\rho'),
\end{equation*}
and introduce the radial Green's function $g_l$ subject to the
equation
\begin{equation}
\ddot g_l + \frac{2 r'}r \dot g_l - \frac{l(l+1)}{r^2}
g_l =- \frac{\delta (\rho - \rho')}{r^2}.\label{maineq}
\end{equation}
We represent the solution of this equation in the following form
\begin{equation}
g_l = \theta(\rho-\rho') \Psi_1(\rho) \Psi_2 (\rho') +
\theta(\rho'-\rho) \Psi_1(\rho') \Psi_2 (\rho), \label{radialform}
\end{equation}
where functions $\Psi$ are the solutions of the corresponding
homogeneous equation
\begin{equation}
\ddot \Psi + \frac{2 r'}r \dot \Psi - \frac{l(l+1)}{r^2} \Psi =0,\label{radial}
\end{equation}
satisfying the boundary conditions
\begin{equation}
\lim_{\rho \to +\infty}\Psi_1 = 0,\ \lim_{\rho \to +\infty}\Psi_2 =
\infty.\label{condlimit}
\end{equation}
If one substitutes (\ref{radialform}) to (\ref{maineq}) the condition
for the Wronskian emerges:
\begin{equation} 
W(\Psi_1,\Psi_2) = \Psi_1 \dot \Psi_2 - \dot \Psi_1 \Psi_2 =
\frac 1{r^2(\rho)}.\label{wronskian1}
\end{equation}

In the case of scalar massive particle with scalar field with non-minimal
coupling we have different radial equation
\begin{equation}
g_l'' + \frac{2r'}{r}g_l' - \left(m^2 + \frac{l(l+1)}{r^2} + \xi
R\right)g_l = -\frac{\delta(\rho-\rho')}{r^2}.
 \end{equation}

We consider the radial equation in domains $\rho > 0$ and $\rho <
0$ and obtain a pair of independent solutions $\phi^1,\phi^2$ for
each of the domains separately. We do not need to consider two
domains if it is possible to construct solutions that are
$C^1$-smooth over all space. However, this is not the case for
many situations. For the two kinds of the throat profile
considered below we may easily construct solutions for the two
domains separately (but not for all space). After that a procedure
developed here allows to construct $C^1$-smooth solution over all
space.

Let us consider in detail the simple case of the symmetric throat
profile: $r(-\rho) = r(\rho)$. In this case we obtain the radial Green function 

\bs\label{g_lGen3}
1. $\rho>\rho'>0$
\begin{equation}
\label{g^1Gen3}
g_l^{(1)}(\rho,\rho') =  -\frac 1{A_+}\phi^2_+(\rho')\phi^1_+(\rho) +
\left.\frac 1{A_+}\frac{W_+(\phi^1_+,\phi^2_+)}{W_+(\phi^2_+,\phi^2_+)}
\right|_0 \phi^2_+(\rho')\phi^2_+(\rho)
\end{equation}

2. $\rho < \rho'$ and $\rho'>0,\ \rho<0$
\begin{equation}
g_l^{(2)}(\rho,\rho') = -\left.\frac 
1{A_+}\frac{W(\phi^1_+,\phi^2_+)}{W_+(\phi^2_+,\phi^2_+)}
\right|_0\phi^2_+(\rho')\phi^2_+(-\rho)
\end{equation}
\es
Here $W_+(y_1,y_2)=y_1\dot y_2 + \dot y_1 y_2$ and $A_\pm =
W(\phi^1_\pm,\phi^2_\pm)r^2(\rho)$.  

\begin{figure}[pb]
\centerline{\includegraphics[scale=0.5]{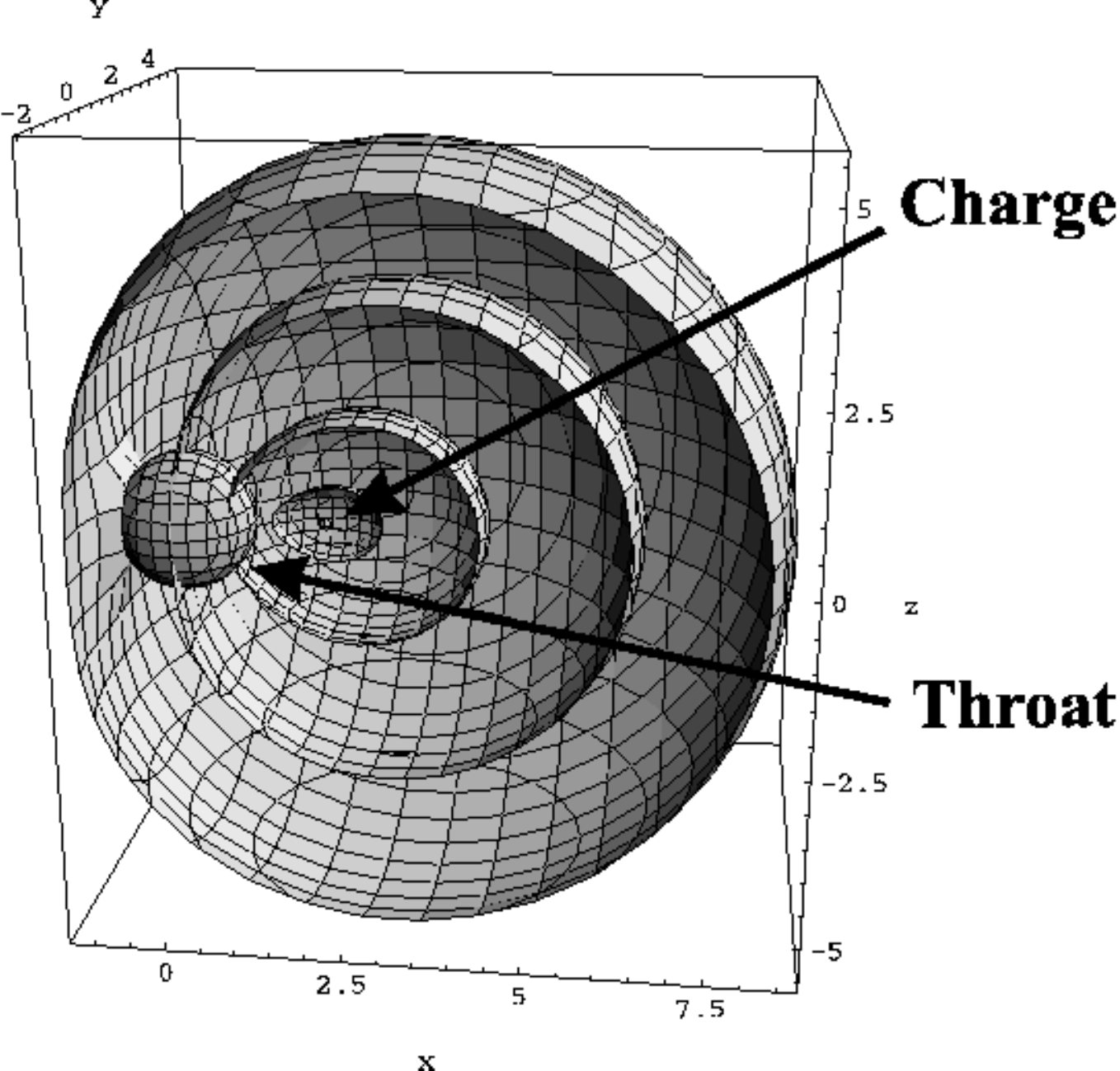}}
\vspace*{8pt}
\caption{Several surfaces of constant potential are shown. The
charge $e=1$ is at the point $x=2,y=z=0$. The small sphere is the
throat of the wormhole, $\rho=0$. We observe that some surfaces go
under the throat to another universe.} \label{Fig:p}
\end{figure}

\subsection{Profile $r = a + |\rho|$}

i) Electromagnetic field. The Green function reads\cite{KhuBah}
\begin{eqnarray*}
4\pi G^{(1)}(x;x') &=& \frac 1{\sqrt{r(\rho)^2
-2r(\rho)r(\rho')\cos\gamma + r(\rho')^2}}\adb\\
&-& \frac{1}{2a}\ln\left|1+\frac{2t}{1-t+\sqrt{t^2 -
  2t\cos\gamma+1}}\right|,\adb\\
4\pi G^{(3)}(x;x') &=& \frac{t}{a\sqrt{t^2 - 2t\cos\gamma+1}} -
  \frac{1}{2a}\ln\left|1+\frac{2t}{1-t+\sqrt{t^2 -
  2t\cos\gamma+1}}\right|,
\end{eqnarray*}
where $t = \frac{a^2}{r(\rho) r(\rho')}$. The surfaces of constant energy are
shown in Fig. \ref{Fig:p}. 

The the self-potential is
\begin{equation}\label{U}
U=\frac{e^2}{4 a} \ln [1-\frac{a^2}{(a+\rho)^2}].
\end{equation}
The self-force
\begin{equation*}
\mathbf{F} = -\nabla U
\end{equation*}
has the radial component only
\begin{equation*}
F^\rho = -\partial_\rho U = -\frac{ae^2}{2r(\rho)^3}\frac
1{1-\frac{a^2}{r(\rho)^2}}.
\end{equation*}
The self-force is always attractive, it turns into infinity at the
throat and goes down monotonically to zero as $\rho \to \infty$.
We may compare this expression with its analog for Schwarzschild
space-time with Schwarzschild radius $r_s = a$:
\begin{equation*}
F^r = +\frac{ae^2}{2r^3}\sqrt{1-\frac{a^2}{r^2}}.
\end{equation*}
Important observations are:

1) The self-force in the wormhole space-time has an opposite sign
-- it is attractive.

2) Far from the wormhole throat and from the black hole we have
the same results but with opposite signs
\begin{eqnarray*}
F^\rho_{wh} &=& -\frac{ae^2}{2\rho^3},\\
F^\rho_{bh} &=& +\frac{ae^2}{2\rho^3}.
\end{eqnarray*}

3) At the Schwarzschild radius $r_s = a$ the self-force equals
zero, whereas at the wormhole throat it tends to infinity. The
latter discrepancy originates in the selected throat profile
function that leads to the curvature singularity at the throat.

ii) Scalar field. For massless particle the self-potential read\cite{BezKhu} 
\begin{equation}
U(\rho) = - \frac{ae^2(1-8\xi)}{4
r^2} \Phi\left(\frac{a^2}{r^2},1,1-4\xi\right),\label{UPhi}
\end{equation}
where
\begin{equation}
 \Phi\left(\frac{a^2}{r^2},1,1-4\xi\right) = \sum_{n=0}^\infty (1-4\xi+n)^{-1}
\left(\frac ar\right)^{2n}.
\end{equation}
For massive particle we obtain 	for self-energy the following
expression
\begin{equation}
 U(\rho) = -e^2 \sum_{l=0}^\infty \nu
\left.\frac{m a(I_\nu K_\nu' + I_\nu'K_\nu) +
(8\xi-1) I_\nu K_\nu}{2m a K_\nu K_\nu' + (8\xi-1)
K_\nu^2}\right|_{m a} \frac{K_\nu^2(m r)}{r},
\end{equation}
where $I_\nu$ and $K_\nu$ are the Bessel function of the secong kind. 

\begin{figure}[pb]
\centerline{\includegraphics[scale=0.7]{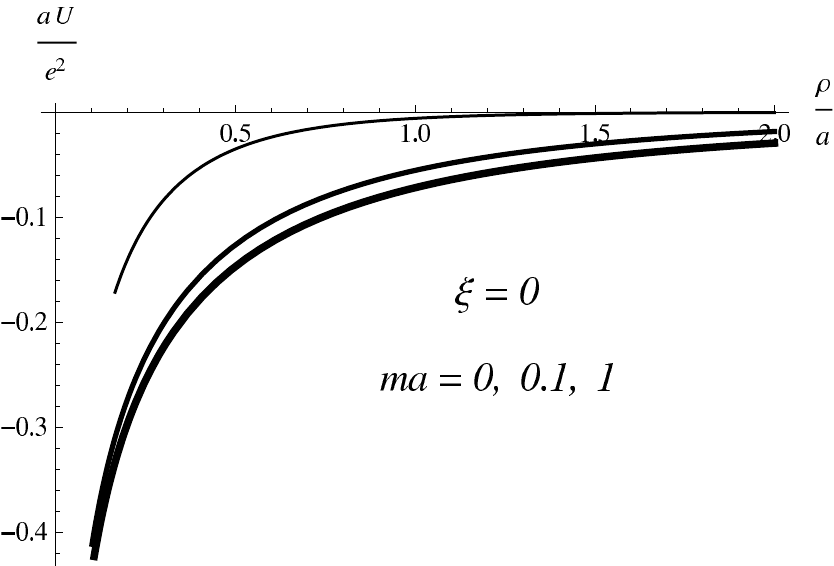}\includegraphics[scale=0.7]{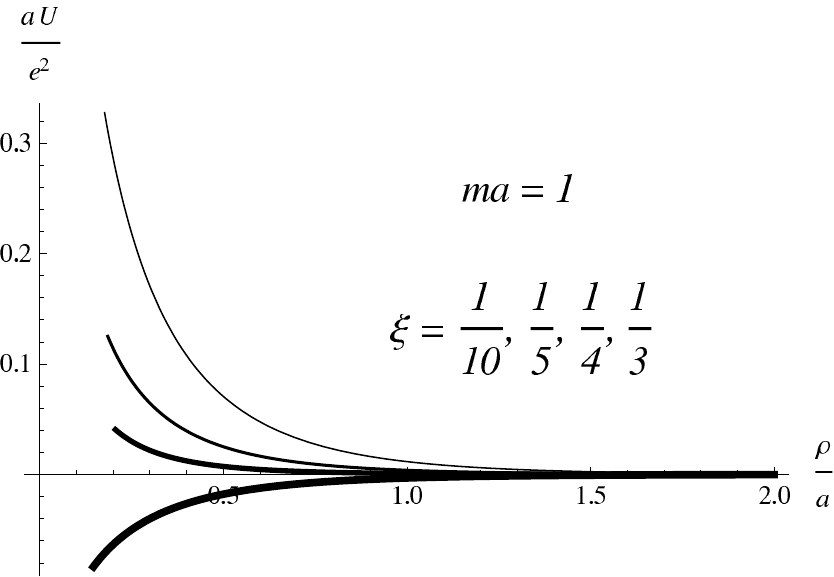}}
\vspace*{8pt}
\caption{The numerical simulation of the self-energy of a massive
scalar field for $\xi =0$ and for different parameters $ma=0$ (thick line), 
$ma=0.1$ (middle thickness) and $ma = 1$ (thin line). In the figure at right we
show the numerical simulation for $ma=1$ and for different parameters $\xi
=1/10$ (thick line) up to $\xi =1/3$ (thin line).}\label{fig:1}
\end{figure}

The numerical simulations of the self-energy are shown
in Fig. \ref{fig:1}. 
We note that the massive field will produce a self-force which is
localized close to the throat. It falls down exponentially fast as $e^{-mr}$ far
from the throat. This behavior is in agreement with Linet result.\cite{Linet:1986:wescs}

\begin{figure}[pb]
\centerline{\includegraphics[scale=0.7]{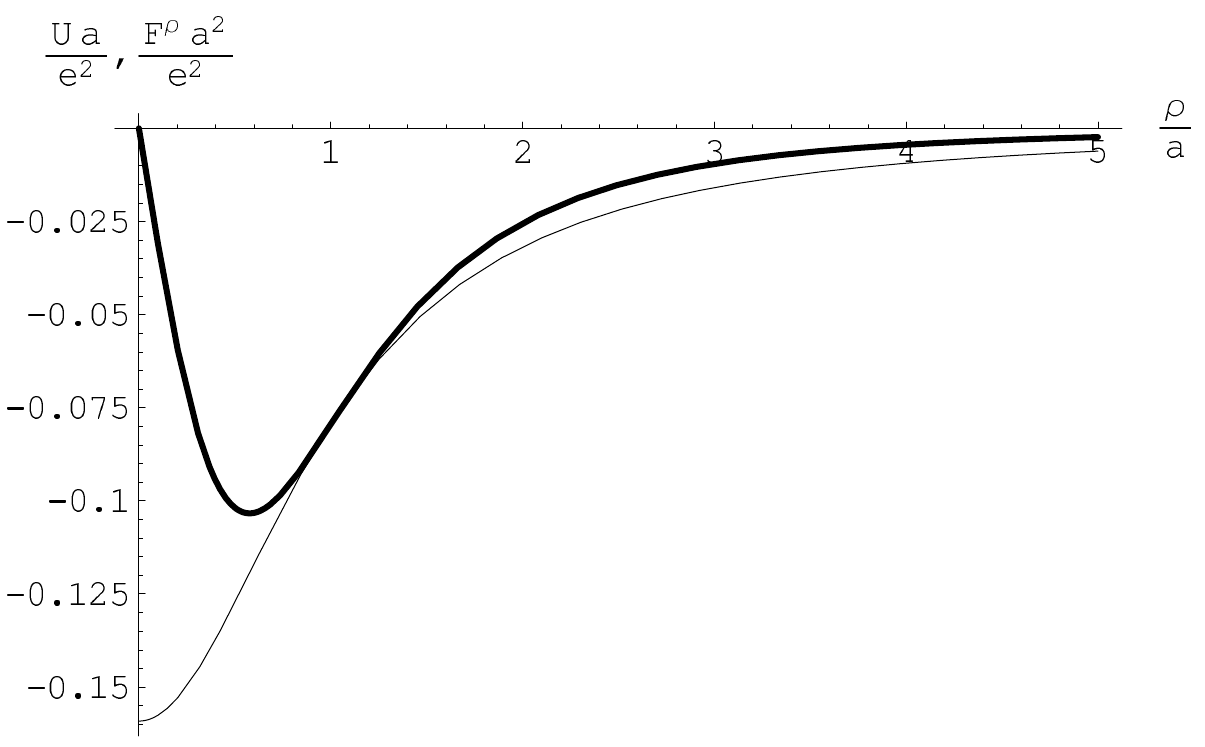}}
\vspace*{8pt}
\caption{Thin line is plot of potential and
thick line is a plot of the self-force (\ref{U2}). The force has an extreme at
point $\rho^* = a/\sqrt{3}$.} \label{Fig:pf}
\end{figure}

\begin{figure}[pb]
\centerline{\includegraphics[scale=0.9]{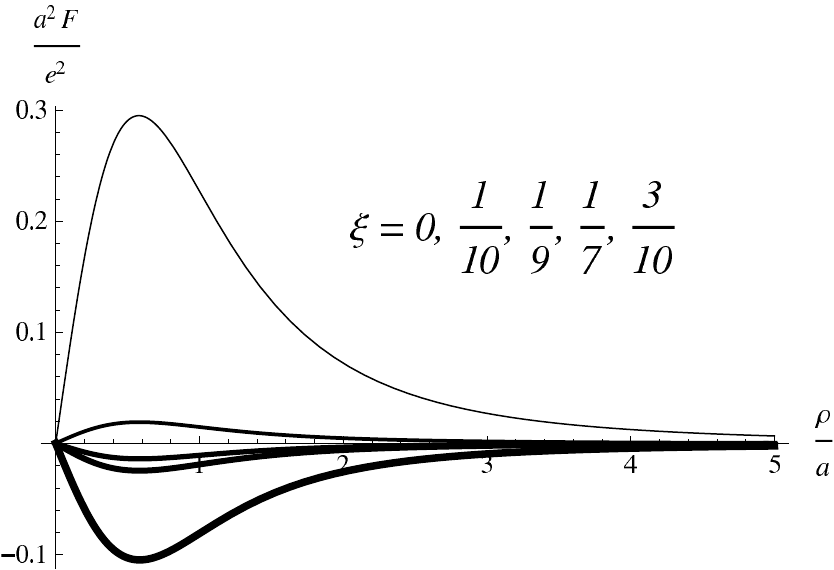}}
\vspace*{8pt}
\caption{The numerical simulation of the
self-force on a massless scalar field for profile $r=\sqrt{\rho^2 + a^2}$ for
different parameters from $\xi=0$ (thick line) up to $\xi = \frac{3}{10}$ (thin
line). For $\xi = \frac 18$ it is zero.}\label{fig:w}
\end{figure}

\subsection{Profile $r = \sqrt{a^2+\rho^2}$}

i) Electromagnetic field. The self-energy and self-force read\cite{KhuBah}
\begin{equation}
U = -\frac{e^2}{2\pi} \frac a{\rho^2 +a^2}, \ F^\rho = \partial_\rho U =
-\frac{e^2}{\pi}
\frac{a\rho}{(\rho^2 +a^2)^2}.\label{U2}
\end{equation}
The self-force is everywhere finite and equals zero at the throat. Far from the
wormhole we have
\begin{equation*}
F^\rho \approx
-\frac{e^2}{\pi} \frac{a}{\rho^3}.
\end{equation*}
Thus the self-force is always attractive. It has maximum value at
distance $\rho^*=a/\sqrt{3}$ with magnitude $F^\rho_{max} =
3\sqrt{3}e^2/16\pi a^2$. The plots of the potential and the
self-force are shown in the Fig. \ref{Fig:pf}.

ii) Scalar field. Massless case. The expression for this case was obtained in
Ref.~\refcite{BezKhu} and it has the following form ($\mu = \sqrt{2\xi}$) 
\begin{equation}
\label{U_renorm'd}
U(\rho) = \frac{e^2}{2} \left[- \frac 1r + \frac 1r \sum_{k=1}^\infty
\zeta_H(2k,\frac 32) j_{2k}^s  + \frac{\cos(2\mu \arctan \frac\rho a) -
\cos (\pi\mu)}{2 a \mu
\sin\pi\mu}\right],
\end{equation}
where
\begin{eqnarray*}
j_2^s  &=& - \zeta\frac{-1+ \dot{r}^2 + 2r \ddot{r}}{8},\adb\\
j_4^s  &=& \frac{3\zeta ^2}{128}  \left(\dot{r}^2+2 r
   \ddot{r} - 1\right)^2 - \frac{r \zeta}{16}   \left(2 \ddot{r}
   \dot{r}^2 + 4 r r^{(3)} \dot{r} + r \left(2
   \ddot{r}^2+r r^{(4)}\right)\right).
\end{eqnarray*}
As expected it is zero for $\xi = 1/8$ and it is
divergent for $\xi = 1/2$. Far from the throat we obtain
\begin{equation}
 U \approx -\frac{e^2}{2\rho^2} \frac{a\mu}{\tan \pi\mu}.\label{Ufar}
\end{equation}
The numerical
simulations are reproduced in Fig. \ref{fig:w}.

For massive case the self-force is localized close to the throat of the
wormhole.

\subsection{General Profile}\label{Sec:4}

i) Electromagnetic field. To consider general profile of the throat let us
perform the WKB analysis of radial equation with WKB parameter $\nu = l+1/2$. We
arrive with the following formula\cite{KhuBah}
\begin{equation}
U(\rho) = \frac{e^2}{2} \left[- \frac 1r + \frac 1r \sum_{k=1}^\infty
\zeta_H(2k,\frac 32) j_{2k}^e + \frac 1a \varphi^2_+(\rho) -
\frac 1{2a} \varphi^2_+(\rho)\frac{\varphi^2_+(\rho)}{\varphi^2_+(0)}\right],
\end{equation}
where the solutions for zero mode read
\begin{equation}
\varphi_+^1 = 1,\ \varphi_+^2 = \int_\rho^\infty \frac{a}{r^2}d\rho,
\label{phizero}
\end{equation}
and 
\begin{eqnarray*}
j_2^e &=& - \frac{-1+\dot r^2 + 2r\ddot r}{8},\adb\\
j_4^e &=& \frac 1{128} [3+3\dot r^4 - 12 r\ddot r - 4r^2
\ddot r^2 -2 \dot r^2 (3+2r\ddot r) - 32 r^2 \dot r r^{(3)} - 8 r^3
r^{(4)}].
\end{eqnarray*}

Analysis of this expression for great distance from the throat of the
wormhole gives the following expression for the self-potential
\begin{equation*}
U = -\frac {e^2}{4\rho^2}\frac a{\varphi^2_+(0)} = -\frac {e^2}{4\rho^2}
\left[\int_0^\infty \frac{d\rho}{r^2(\rho)}\right]^{-1}.
\end{equation*}
Note that it is always negative, hence the self-force is an
attractive force. All information about the specific throat
profile is encoded in the factor
\begin{equation*}
\int_0^\infty \frac{d\rho}{r^2(\rho)}.
\end{equation*}

ii) Scalar field. Massless case. For arbitrary profile of the wormhole we have
the following formula
\begin{equation}
U(\rho) = \frac{e^2}{2} \left[- \frac 1r + \frac 1r \sum_{k=1}^\infty
\zeta_H(2k,\frac 32) j_{2k} (\rho,\rho)+ g_0(\rho)\right],\label{Ugeneral}
\end{equation}
with the same $j_{2k}^s$ as above and  
\begin{eqnarray}
 g_0 &=&-\frac 1{A_+}\varphi^2_+(\rho)\varphi^1_+(\rho) + \frac
1{2A_+} \left(
\frac{\varphi^1_+}{\varphi^2_+} +
\frac{\varphi'^1_+}{\varphi'^2_+}\right)_0
\varphi^2_+(\rho)\varphi^2_+(\rho),\label{ZeroModeGeneral}
\end{eqnarray}
where $A_+= W_+(\varphi^1_+,\varphi^2_+)r^2(\rho)$. The functions
$\varphi^{1,2}_+$ are the solutions of the equation
\begin{equation}
\varphi'' + \frac{2 r'}r \varphi'  - \xi R \varphi = 0.
\end{equation}
Unfortunately, differently from the electromagnetic field case, there is no
general solution of this equation for arbitrary $\xi$ and $r$. Far from the
throat we obtain the following expression 
\begin{equation}
 U \approx -\frac{e^2}{2\rho^2} A \label{uinf}.
\end{equation}
But we can not make any conclusion about sign of these expression because the
constant $A$ is expressed in terms of the zero mode which can not be found in
closed form for arbitrary profile of the throat.

\section{The Massive Wormhole Space-Time}

The line element of massive wormhole has the following form\cite{Bron73,Ell73}
\begin{equation}\label{metric2}
ds^2 = -e^{-\alpha(\rho)} dt^2 + e^{\alpha(\rho)}d\rho^2 + r^2(\rho) d\Omega^2,
\end{equation}
where
\begin{eqnarray}
r^2(\rho) &=& (\rho^2 + n^2 -m^2) e^{\alpha(\rho)},\\
\alpha(\rho) &=& \frac{2m}{\sqrt{n^2-m^2}} \left\{ \frac{\pi}{2}
- \arctan \left(\frac{\rho}{\sqrt{n^2-m^2}}\right) \right\}.
\end{eqnarray}
The radial coordinate $\rho$ may be positive as
well as negative, too.  The square of the sphere of radial coordinate $\rho$,
$S=4\pi r^2(\rho)$, is minimized for $\rho = m$.

The renormalization procedure for the space-time with line element
(\ref{metric2}) is not so simple as it was for massless wormhole.\cite{KhuBah,BezKhu} The point is that the space-time under consideration has no ultrastatic form, that is $g_{tt} \not = 1$. For this reason the
equation for $A_t$ in static case does not coincide with that for
scalar $3D$ Green function and we can not use the standard
formulas for DeWitt-Schwinger expansion of the $3D$ Green
function. The renormalization procedure for this case was developed in
Ref.~\refcite{KhuPopLip}. The singular part of potential which has to be
subtracted has the following form
\begin{equation} \label{GDS}
A_{(t)}^{sing}(x^i; {x^i}') = -e \left(\frac{1}{\sqrt{2 \sigma}} +
\frac{g_{t't',i'}\sigma^{i'}}{4{g_{t't'}} \sqrt{2 \sigma}} \right),
\end{equation}
where $\sigma$ is one-halfe  of square of geodesic distance. 

Taking into account this singular part we obtain the following form of the 
self-potential
\begin{equation}\label{UselfWH}
U^{self} = -\frac{e^2}{\rho^2 + n^2 - m^2} \frac{m
e^{-\alpha}}{2\tanh\pi b}
\end{equation}
and tetrad component of the self-force
\begin{equation}\label{FselfWH}
{\cal F}^{(\rho)} = -\partial_\rho U^{self} = \frac{e^2}{(\rho^2 + n^2 - m^2)^2}
  \frac{m (m-\rho)e^{-\alpha}}{\tanh\pi b}.
\end{equation}

\begin{figure}[pb]
\centerline{\includegraphics[scale=0.7]{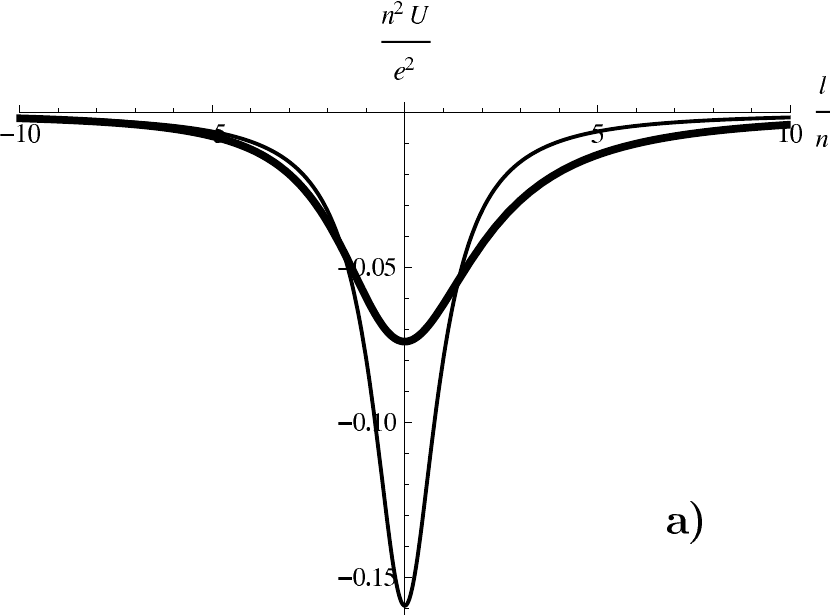}\includegraphics[scale=0.7]{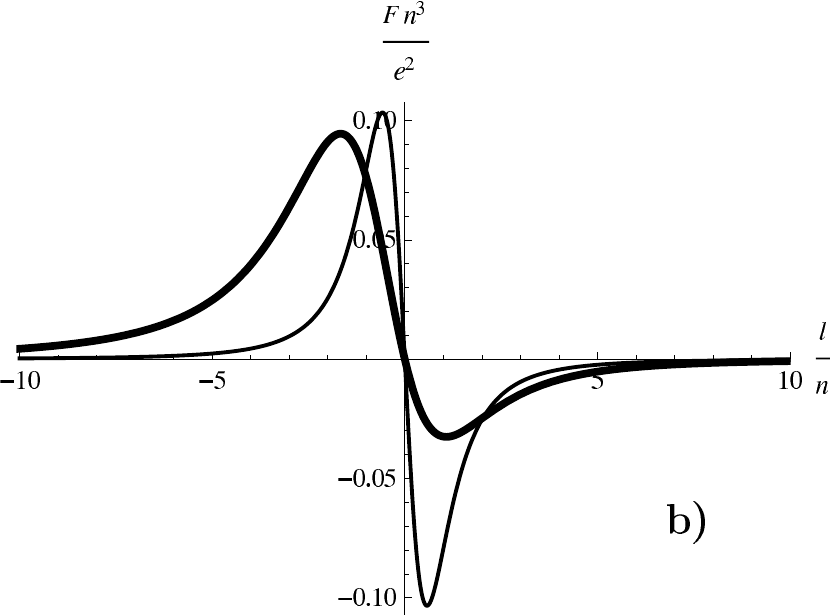}}
\vspace*{8pt}
\caption{The self-energy (a) and the self-force (b) for massless case (thin
curves)  and for massive wormhole case (thick curves) for $m/n=0.7$. The zero
value of the $l$ corresponds to the sphere of the minimal
square in both cases.}\label{fig:self}
\end{figure}

For massless wormhole, $m\to 0$, we recover results obtained in Ref.
\refcite{KhuBah}. Far from the wormhole's throat we obtain
\begin{equation}\label{ratios}
\frac{U_{\rho\to +\infty}^{self}}{U_{\rho\to -\infty}^{self}} = \left|
\frac{{\cal F}^{(\rho)}_{\rho\to +\infty}}{{\cal F}^{(\rho)}_{\rho\to
-\infty}}\right| = e^{2\pi b},
\end{equation}
which is the consequence of the non symmetric form of the space-time under
consideration. The numerical calculations are shown in Fig. \ref{fig:self}.

\section{Conclusion}

In above sections we considered the self-energy and self-force for particle
with electric or scalar charges in space-time of the wormholes. The general
conclusion for electrically charged particle is that they will be attracted to
wormhole throat for arbitrary profile of the throat and arbitrary mass of the
wormhole. The mass of the wormhole slightly changes the symmetric form of the
self-force. This is due to the fact that the massive wormhole space-time has
asymmetric form. From the astrophysical point of view it means that the
wormhole's throat must be surrounded by particles. The self-energy of scalar
particle exponentially falls down at the distance of the Compton wavelength of
scalar filed and falls down polynomial for massless field. The sign of the
self-force depends on the non-minimal coupling of the scalar field. 

\section*{Acknowledgments} 
This work was supported by the Russian Foundation for Basic Research Grant No.
11-02-01162-a.


\begin{thebibliography}{99}
\bibitem{EinRos35} A. Einstein and N. Rosen, {\it Phys. Rev.}, {\bf 48}
73 (1935).
\bibitem{WheBook} J. A. Wheeler, {\it Phys. Rev}, {\bf 97}, 511 (1955).
\bibitem{MorTho88} M. S. Morris and K. S. Thorne, {\it Am. J. Phys.}, {\bf 56} 
395 (1988).
\bibitem{MorThoYur88} M. S. Morris, K. S. Thorne and U. Yurtsever, {\it Phys.
Rev. Lett.}, {\bf 61} 1446 (1988).
\bibitem{VisBook} M. Visser, {\it Lorentzian Wormholes: From Einstein to
Hawking}, (American Institute of Physics, Woodbury, NY, 1995).
\bibitem{Kra00} S. Krasnikov, {\it Phys. Rev. D}, {\bf 62} 084028 (2000).
\bibitem{KhuSus02} N. R. Khusnutdinov and S. V. Sushkov, {\it Phys. Rev. D}, 
\textbf{65} 084028 (2002); N. R. Khusnutdinov, {\it Phys. Rev. D}, {\bf 67}
124020 (2003).
\bibitem{Gar05} R. Garattini, {\it Class. Quant. Grav.}, {\bf 22} 1105
(2005).
\bibitem{Arm02} C. Armendariz-Picon, {\it Phys. Rev. D}, {\bf 65} 104010
(2002).
\bibitem{Sus05} S. Sushkov, {\it Phys. Rev. D}, {\bf 71} 043520 (2005).
\bibitem{Lob05} F. S. N. Lobo, {\it Phys. Rev. D}, {\bf 71} 084011 (2005).
\bibitem{Lob05_2} F. S. N. Lobo, {\it Phys. Rev. D}, {\bf 71} 124022 (2005).
\bibitem{DamSol07} T. Damour, S. N. Solodukhin, {\it Phys. Rev. D}, {\bf 76}
024016 (2007).
\bibitem{Nar05} R. Narayan, {\it New J. Phys.}, {\bf 7} 199 (2005).
\bibitem{KarNovSha06} N. S. Kardashev, I. D. Novikov and A. A. Shatskiy,
{\it Int. J. Mod. Phys. D}, {\bf 16} 909 (2007).
\bibitem{DeWBre60} B. S. DeWitt, R. W. Brehme, {\it Ann. Phys.}, {\bf 9} 220
(1960).
\bibitem{Poi} E. Poisson,  {\it Living Reviews in Relativity}, {\bf 7} 6
(2004).
\bibitem{Khu05} N. R. Khusnutdinov, {\it Physics - Uspekhi}, {\bf 48} 577
(2005)[{\it Uspekhi Fizicheskikh Nauk}, {\bf 175} 603 (2005)].
\bibitem{Lin86} B. Linet, {\it Phys. Rev. D}, {\bf 33} 1833 (1986).
\bibitem{Gal90} D. V. Gal'tsov, {\it Fortschr. Phys.}, {\bf 38} 945 (1990).
\bibitem{KhuBez01} N. R. Khusnutdinov and V. B. Bezerra, {\it Phys. Rev. D},
{\bf 64} 083506 (2001).
\bibitem{KhaKhuSus06} A. R. Khabibullin, N. R. Khusnutdinov and
S. V. Sushkov, {\it Class. Quantum Grav.}, {\bf 23} 627 (2006).
\bibitem{Bron73} K. A. Bronnikov, {\it Acta Phys. Pol. B}, {\bf 4} 251 (1973).
\bibitem{Ell73} H. G. Ellis, {\it J. Math. Phys.}, {\bf 14} 104 (1973).
\bibitem{BezKhu} V. B. Bezerra and N. R. Khusnutdinov, {\it Phys. Rev. D}, {\bf
79} 064012 (2009). 
\bibitem{KhuBah} N. R. Khusnutdinov and I. V. Bakhmatov, {\it Phys.Rev. D}, {\bf
76} 124015 (2007).
\bibitem{Linet:1986:wescs} B. Linet, {\it Ann. Inst. Henri Poincare A}, {\bf
45} 249 (1986).
\bibitem{KhuPopLip} N. R. Khusnutdinov, A. A. Popov and L. N. Lipatova, {\it
Class. Quantum Grav.}, {\bf 27} 215012 (2010).
\end{thebibliography}
\end{document}